\documentclass{PoS}

\newcommand{\HI}{H{\sc i}}

\title{From Feast To Famine: the Role of HI in Group Evolution}

\ShortTitle{From Feast To Famine: the Role of HI in Group Evolution}

\author{\speaker{Michelle Cluver}%
        \thanks{MC would like to acknowledge funding from the National Research Foundation as part of the Research Career Advancement Programme.}\\
       University of the Western Cape\\
       E-mail: \email{michelle.cluver@gmail.com}}

\author{Lourdes Verdes-Montenegro\\
       Instituto de Astrof\'{i}sica de Andaluc\'{i}a (IAA)}
       
\author{Kelley Hess\\
       Kapteyn Astronomical Institute, University of Groningen}       

\abstract{Galaxies in the local universe are most commonly found in groups and are thought to be ``pre-processed" in this environment before being consumed by clusters. Yet we know very little about the gastrophysics of these systems, how they evolve and how this environment is connected to the quenching of star-forming galaxies. In particular, the role of intragroup gas has been challenging to uncover due to observational constraints and the limitations of radio telescopes to date. Sensitive, interferometric \HI\ observations of galaxy groups, combined with multiwavelength tracers of stellar mass, star formation and shocks, is necessary to examine the physical processes transforming galaxies from star-forming to quenched. These laboratories may be key to understanding the dominant mechanisms driving galaxy evolution. MeerKAT uniquely combines a large field of view, column density sensitivity, and excellent UV coverage on short baselines ensuring sensitivity to diffuse gas. This design makes it a compelling instrument for the study of intragroup and circumgroup gas, quenching in galaxy groups, and for tracing evolutionary pathways within the group environment.}

\FullConference{MeerKAT Science: On the Pathway to the SKA\\
                 25-27 May, 2016\\
                 Stellenbosch, South Africa}

\begin{document}

\section{Introduction}

The fundamental principle underlying our entire Universe is that neutral hydrogen gas will collapse under gravity to form the stars that build a galaxy. It is the physics that drives everything, but this simple description can be deceptive. The details of how star formation proceeds entails a number of, often competing, processes (see e.g. Schiminovich et al. 2007). Feedback, notably when star formation is particularly vigorous or due to the intense feeding of a supermassive black hole, gas accretion, and the effects of secular (slow and steady) evolution are difficult to disentangle. An additional complication is that galaxies rarely live alone. The primordial fluctuations of the Early Universe, seen as minute temperature differences in the Cosmic Microwave Background, spawned a Cosmic Web where galaxies congregate in groups, clusters and, on larger scales, filaments, walls and superclusters. 

Large galaxy surveys have revealed that galaxies in the local universe show a strong bimodality, separating into two main populations when viewed in a colour-magnitude diagram: a ``blue cloud" of star-forming galaxies and a ``red sequence" of quenched, passively evolving systems (Strateva et al. 2001). The key question is therefore, what are the mechanisms that cause a galaxy to transition from a gas-rich, star-forming disk (feast) to a gas-poor ``red and dead" galaxy (famine)? Since not all galaxies will follow the same evolutionary pathway, our task is to determine which paths are traced most often, and which mechanisms dominate this process.

The role of environment in this process has been somewhat contentious in the field. Observations suggest that the density of a galaxy's environment effects its properties (e.g. Butcher \& Oemler 1978, Dressler 1980, Blanton \& Moustakas 2009). However, Peng et al. (2010) suggest that environmental quenching only plays a role as large scale structure develops, while ``mass quenching" affects galaxies around and above M$^\star$ and is ubiquitous. However, the physical mechanism that drives mass quenching is not known.  

Schawinski et al. (2014) combined SDSS, {\it GALEX} and Galaxy Zoo to study quenching of late- and early-types. They find that late-type quenching appears consistent with a gas supply that is severed, followed by slow secular evolution to the red sequence. The quenching of early-type galaxies, however, is rapid and suggests the sudden destruction of the gas reservoir.

Disk galaxies can self-regulate their star formation (e.g. Ostriker \& Shetty 2011). But just one grazing encounter between two galaxies can produce significant changes in one or both systems (Toomre \& Toomre 1972). Repeated high velocity ``fly-by" interactions between multiple galaxies, as would be common in groups and clusters, is termed ``galaxy harassment". This can have a dramatic effect on a galaxy; bursts of star formation, asymmetries, changes in morphology, warps, and the formation of tidal tails are all evidence of immense disruption to a so-called ``closed-box" system. The fraction of systems that go on to merge are an additional consideration, and potentially a major driver of star formation in the early Universe.

The group environment is key since most galaxies in the Universe are expected to reside in groups, and in the paradigm of hierarchical structure formation groups coalesce to form clusters of galaxies. In other words, galaxies are ``pre-processed" by the group environment before they are assimilated into clusters. In groups, tidal interactions dominate and systems are relatively simple, as compared to clusters where galaxies in general are more evolved and where the hot X-ray-emitting intracluster medium is an additional complication. Tidal torques funnel gas to the centre, encouraging the growth of a bulge, while tidal stripping continues. Bekki \& Couch (2011) used chemodynamical simulations to show that spirals can transform efficiently into early-type sprials (S0's) in a group environment via repetitive slow encounters. This process also forms intragoup cold \HI\ gas. Studying such complex and stochastic interactions relies heavily on simulations and modeling. Although advances in computing make it possible to keep track of innumerable calculations, simulating the effects of gravitational interactions and the resulting torques and tides, reproducing the Universe we see around us remains a challenge. Observations must dictate the prescriptions of physics and chemistry that underpin cosmological simulations.

There is no denying that large surveys have revolutionised our understanding of galaxy evolution. For instance, optical photometry and spectroscopy from the Sloan Digital Sky Survey (York et al. 2000), the impressive ALFALFA \HI\ Survey (Haynes et al. 2011), and the powerful combination of the two (Catinella et al. 2013, Hess and Wilcots 2013) have shown that environmental effects can produce the removal of atomic gas from group galaxies, simultaneously quenching star formation, suggesting a mechanism such as viscous stripping (Nulsen 1982).

Although the statistical analysis driven by large datasets promotes a greater understanding of aggregate evolution, true comprehension lies in the interstellar medium (ISM) physics or ``gastrophysics" of galaxies. Studing the changing metabolism of galaxies provides the crucial prescriptive physics that will advance our progress significantly.

Sensitive observations of the neutral gas in galaxy groups, combined with multiwavelength tracers of stellar mass, star formation and shocks, is necessary to examine the physical processes transforming galaxies from star-forming to quenched. MeerKAT presents a unique opprtunity to study intragroup gas and quenching in galaxy groups. 

\section{Scientific Rationale}

High density environments can cause galaxies to a fall victim to quenching of star formation (e.g. Cortese et al. 2009), often on rapid timescales. Here ram pressure stripping is usually responsible (e.g. Chung et al. 2009). However, similar evidence of accelerated evolution (strong bimodality in mid-infrared colours) was observed in compact groups (Walker et al. 2010, 2012) which in contrast to clusters only weakly X-ray emitting (Rasmussen et al. 2008). Instead they harbour a copious intragroup medium of diffuse neutral gas (Borthakur et al. 2010, 2015) and complex tidal features (Verdes-Montenegro et al. 2001). Compact groups are characterised by high density, lower velocity dispersions compared to clusters ($\sim$ 250\,km\,s$^{-1}$) and relatively shallow gravitational potential wells, which prolongs gravitational interactions. Mid-infrared spectroscopy from {\it Spitzer} of selected compact groups led to the discovery of shock-excited molecular hydrogen being preferentially found in systems undergoing active transformation (Cluver et al. 2010, 2013); that this emission is not linked to active galactic nucleus (AGN) activity suggests a different mechanism accelerating the evolution in these systems.

In Cluver et al. (2013) it was suggested that collisions between galaxies and a 
cold, intragroup component was causing the shock-excitation of H$_2$ (with molecular gas forming due to the collision). The injection of turbulence due to the collision of a disk galaxy with cold gas would disrupt star formation and   viscous stripping (Nulsen 1982) would accelerate the stripping of the gas reservoir available for star formation. Depending on the severity of the collision and the amount of tidal debris involved, galaxies will be affected differently -- some experiencing a rapid shutdown of star formation and others where only outer \HI\ is stripped impacting only the future star formation history of the galaxy.

Given that more general group studies are finding similar evolutionary and environmental behavior (Rasmussen et al. 2012, Catinella et al. 2013, Hess \& Wilcots 2013), the role of \HI\ gas as an agent of star formation suppression (or at the very least a tracer of past interaction) may be a key consideration for semi-analytic galaxy formation models which do not include stripping of the cold interstellar medium. \HI\ which almost uniquely traces the future fuel for star formation, as well as being sensitive to the tidal torques exerted by neighboring galaxies, gives us past and future information about how the system is evolving. With neutral gas observations we can study both how star formation feeding proceeds (using gas kinematics) in systems with no signs of interaction, as well as determine how this reservoir is removed and what quenching systems may be operating in the group environment. An excellent illustration of the power of these observations is shown in Figure 1. The optical image gives no indication of interactions between these galaxies, whereas sensitive \HI\ observations reveal complex evolution.

\begin{figure}[t]
\includegraphics[width=\textwidth]{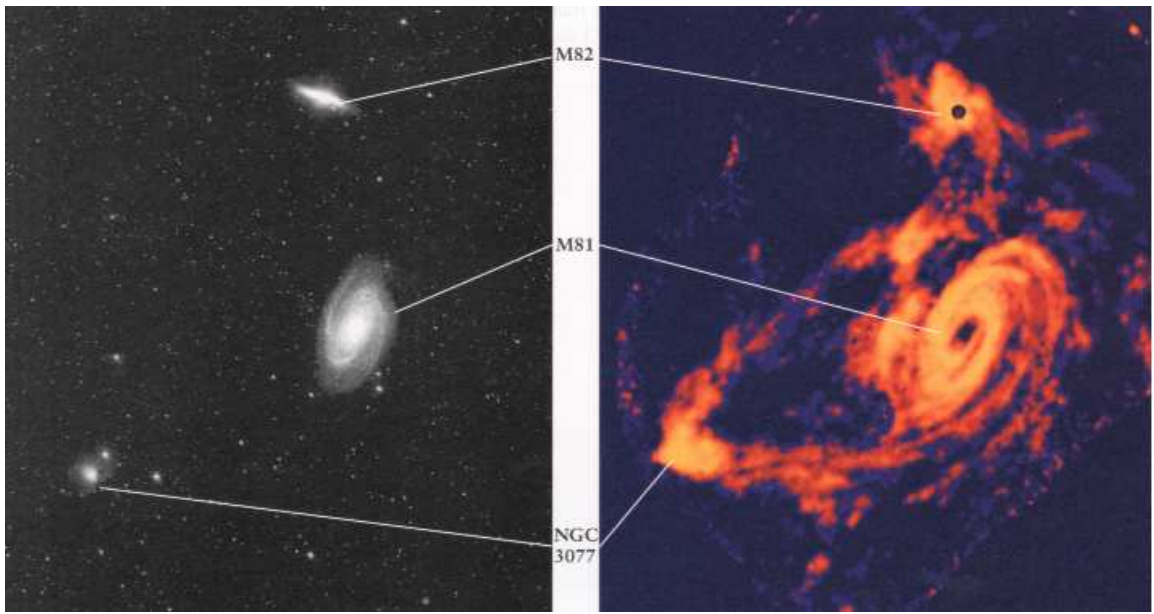}
\caption{Left: Palomar Observatory Sky Survey optical image. Right: High-resolution VLA image of the neutral gas from Yun, Ho \& Lo (1994).}
\end{figure}

To date, sensitive \HI\ observations, combined with spatial resolution (i.e. interferometric observations) and large field of views, have been beyond the limits of modern radio telescopes. However, the SKA pathfinders, MeerKAT and ASKAP will provide an unprecented window on the distribution of neutral gas in the local Universe. This ``missing ingredient" can then be combined with multiwavelength information crucial to forming a complete picture of evolution. Previous work using the GAMA Group Catalogue by Robotham et al. (2013) suggests that the mechanism of environmental suppression is strongly associated with galaxy-galaxy interactions. Neutral gas is the missing and potentially the most important piece of the puzzle. For example, does the intragroup gas ``feed" galaxies via re-accretion of \HI, or conversely, does it disrupt accretion and promote ``starvation"? 

MeerKAT, the South African SKA Pathfinder, will provide sensitive \HI\ measurements over a 1 degree field of view, allowing us to probe the neutral gas fuel in galaxies, as well as intragroup and circumgroup gas. In addition, radio continuum emission will provide star formation, AGN activity and shock measures (in combination with optical spectra). {\bf Deep observations with MeerKAT will allow us to confront the physics and chemistry of group galaxy evolution.}

\section{GAMA Groups}

\begin{figure}[t]
\centering
\includegraphics[width=12cm]{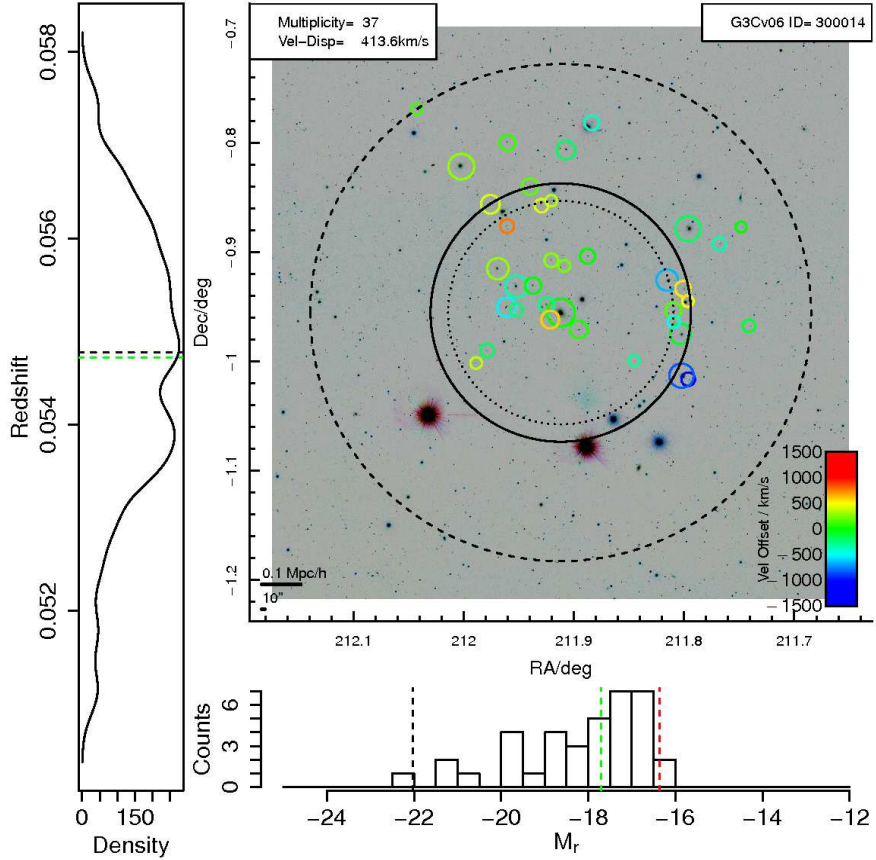}
\caption{Example of a potential MeerKAT group target from the GAMA G3C Group Catalogue (Robotham 2011). Image Credit: A. Robotham.}
\end{figure}

The GAMA dataset (Driver et al. 2009), which combines 21 band multiwavelength imaging (UV, optical, near-, mid-, and far-infrared) with spectroscopic redshifts for $\sim300\, 000$ galaxies (median redshift $\sim$ 0.3) provides state-of-the-art value-added catalogues.  Stellar masses, optical line ratios, S\'{e}rsic profiles etc. provide an unprecedented wealth of data, while the highly complete spectroscopic coverage of the survey has produced a group catalogue superior to that of other large surveys (Robotham et al. 2011). Controlling for stellar mass and taking into account the properties of parent dark matter haloes will be crucial when interpreting results.

{\it WISE} mid-infrared data traces the old stellar population, molecules (polycyclic aromatic hydrocarbons) sensitive to the interstellar radiation field and warm dust. Combined with neutral gas, we can identify if distributed star formation and metallicities are sensitive to the amount of gas, dust and old stars. \HI\ deficient systems (compared to what is expected for a galaxy of similar morphology) will be of particular interest to determine optical excitation mechanisms (shocks, AGN, star formation etc.). 

MeerKAT, the South African SKA Pathfinder, {\bf uniquely} combines column density sensitivity, sensitivity to diffuse emission due to better UV coverage at short baselines, and a large field of view (1 degree). It is therefore better suited to a study of neutral gas in galaxy groups than, for instance, the Very Large Array (VLA) telescope. We aim to obtain MeerKAT observations for at least 15 GAMA groups (using Early Science and Open Time) -- a 12 hour integration on the 64 dish array will achieve a column density sensitivity of 5$\times 10^{18}$ cm$^{-2}$, to our knowledge unprecedented in interferometric group studies to date.  

The focus of the Honours project of Unarine Tshiwawa was to select the best GAMA groups in the equatorial fields ($z < 0.1$) to target with MeerKAT, based on angular size, velocity dispersion and stellar mass properties. One such group is shown in Figure 2. The high velocity dispersion of this group and broad range in stellar mass (10$^{8.6}$ -- 10$^{11.4}$ M$_\odot$) and log specific star formation ($-9.6$ to $-11.7$) make this an ideal target for neutral gas studies.

\section{Hickson Compact Groups}

Verdes-Montenegro et al. (2001) proposed an evolutionary scenario in Hickson Compact Groups (HCGs) where continuous tidal stripping gradually causes galaxies to become \HI-deficient, while the intragroup medium becomes enriched with cold, diffuse gas. This diffuse component was initially missed by the VLA, but at least partly recovered by sensitive Green Bank Telescope (GBT) observations (Borthakur et al. 2010). However, a detailed quantification of \HI\ in HCGs by Verdes-Montenegro et al. (in prep) suggests that a large quantity of cold gas is still unaccounted for. MeerKAT is the ideal telescope to search for this ``missing component". Locating this faint \HI\ is crucial to understanding the mechanisms leading to the accelerated evolution, \HI-deficiency and shock excitaion observed in these systems. The lifetime and fate of this gas may be a key consideration for galaxy evolution models. 

HCGs have been studied extensively and have a wealth of ancillary data that can be combined with radio observations to investigate their evolution. MeerKAT observations of these systems will be highly complementary to observations of the GAMA sample, the MeerKAT survey of Fornax (Serra et al. 2011), and groups observed as part of MIGHTEE (Jarvis et al. 2012).

\section{KAT-7 observes HCG 44}

\begin{figure}[t]
\centering
\includegraphics[width=11cm]{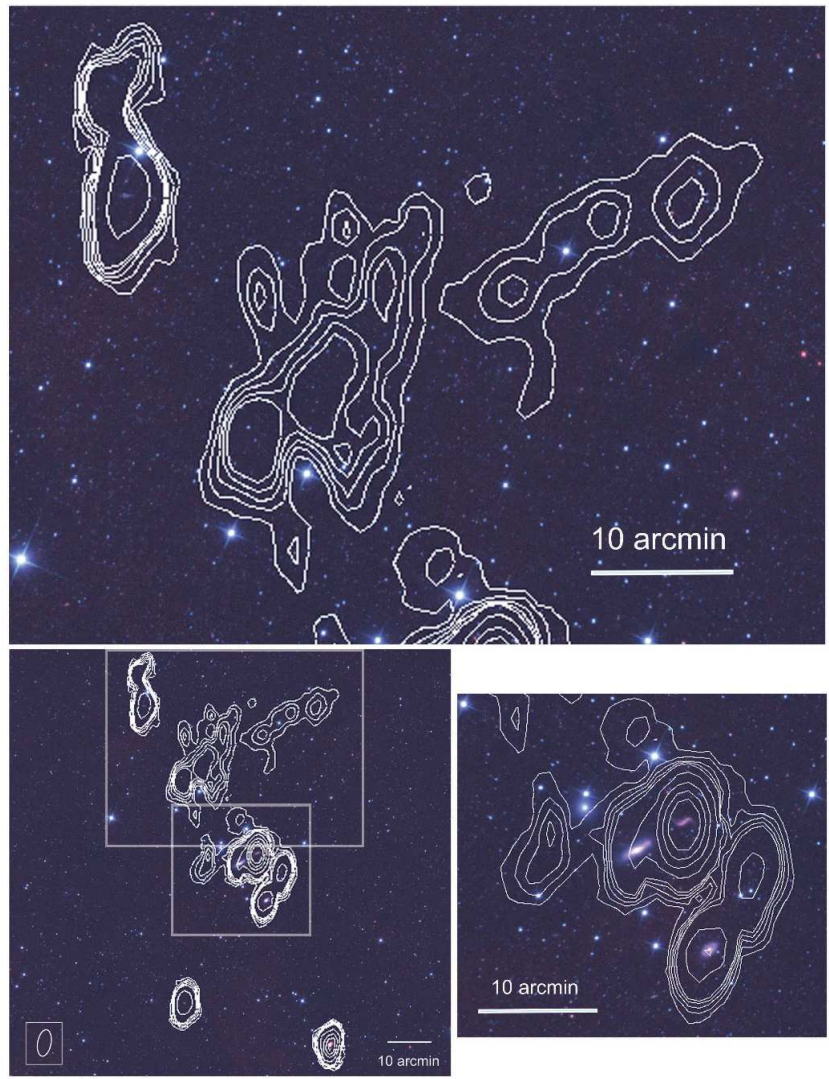}
\caption{A ``drizzled" Wide-field Infrared Survey Explorer (WISE) four-color image of HCG 44 and the extended \HI\ tail: 3.4\,$\mu$m (W1) is blue, 4.6\,$\mu$m (W2) is green, 12\,$\mu$m (W3) is orange, and 22\,$\mu$m (W4) is red. White contours correspond to \HI' column densities of 2, 4, 6, 8, 10, 25, 50, 75, 100 $\times$ $10^{18}$\,cm$^{−2}$ from the KAT-7 observations. Full figure in Hess et al. (2017). }
\end{figure}

As a pilot project of the study of intragroup gas with MeerKAT, we have executed radio observations of HCG 44 using the Karoo Array Telescope, KAT-7, and combined it with data from the Westerbork Sythesis Radio Telescope, WRST (Serra et al. 2013), and from the Arecibo ALFALFA survey (Leisman et al. 2016) to achieve a column density of $<2\times$ $10^{18}$ cm$^{−2}$. This enabled the detection of twice as much gas, extending 33\% further, in the giant \HI\ tail to the north of the group (see Figure 3). These deep observations illustrate the potency of gas stripping within strongly interacting systems, as well as the longevity ($>1$\, Gyr) of such tidal material (Hess, Cluver et al. 2017). 

\section{Tools}

Studing the gas in groups requires being able to separate gas in disks from gas in tails and studying the kinematics of the gas. In interacting systems, this is a challenging task and requires custom tools. The X3D Pathway (Vogt et al. 2016) is a new approach to visualise, share and publish multidimensional datasets using 3D diagrams. The power of this tool is well-illustrated using HCG 91 in Vogt et al. (2015), as shown in Figure 4.

\begin{figure}[t]
\centering
\includegraphics[width=\textwidth]{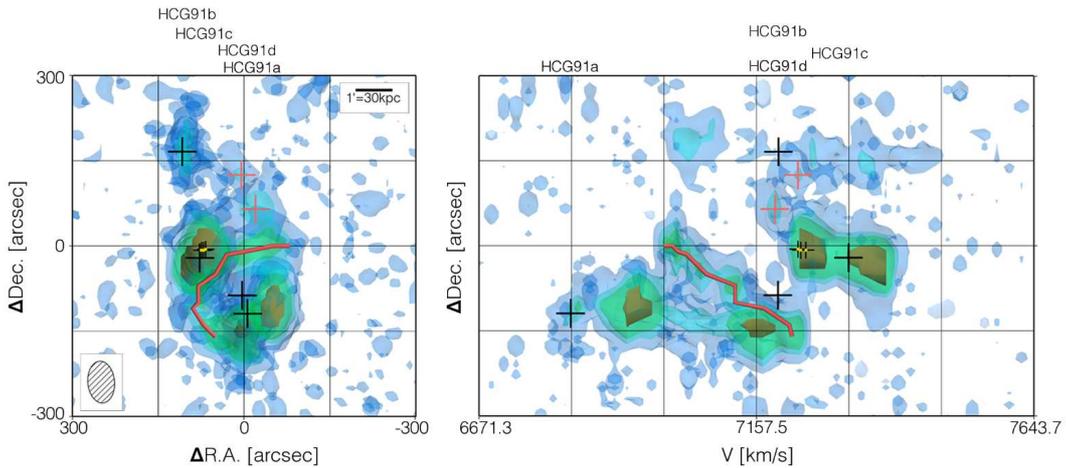}
\caption{The X3D Pathway is a powerful tool for visualising the kinematics and distribution of \HI\ (shown here for VLA data of HCG 91). An interactive version of this image is available in Vogt et al. (2015). }
\end{figure}

\section{Conclusions}

To date, multiwavelength studies of galaxy groups with interferometric \HI\ observations have been largely limited to a relatively small number of nearby, and typically low mass, systems (e.g. Forbes et al. 2006). Conversely, large \HI\ galaxy surveys using single dish observations have been combined with other wavelengths (e.g. ALFALFA+SDSS), but are limited to trends in global properties without spatial neutral gas information (e.g. Catinella et al. 2013), making the mechanisms driving group evolution difficult to disentangle (e.g. Hess \& Wilcots 2013). 

A comprehensive study of the feeding and feedback role of \HI\ in the group environment is now possible using 
ultra-sensitive MeerKAT \HI\ measures with existing multiwavelength data in well-characterised groups, such as provided by GAMA and HCGs.


\begin{thebibliography}{99}


\bibitem{Bek11}Bekki, K., \& Couch, W.~J.\ 2011, {\it MNRAS}, 415, 1783

\bibitem{Bl09}Blanton, M.R. \& Moustakas, J., 2009, {\it ARA\& A}, 47, 159

\bibitem{Bor15}Borthakur, S., Yun, M.~S., Verdes-Montenegro, L., et al.\ 2015, {\it ApJ}, 812, 78

\bibitem{Bor10}Borthakur, S., Yun, M.S., Verdes-Montenegro, L., 2010, {\it ApJ}, 710, 385

\bibitem{BO78}Butcher, H., \& Oemler, A., Jr.\ 1978, {\it ApJ}, 226, 559

\bibitem{Cat13}Catinella, B., Schiminovich, D., Cortese, L., et al.\ 2013, {\it MNRAS}, 436, 34

\bibitem{Ch09}Chung, A., van Gorkom, J.~H., Kenney, J.~D.~P., Crowl, H., \& Vollmer, B.\ 2009, {\it AJ}, 138, 1741

\bibitem{Cl13}Cluver, M.E. et al. 2013, {\it ApJ}, 765, 93

\bibitem{Cl10}Cluver, M.E. et al. 2010, {\it ApJ}, 710, 248

\bibitem{Dres80}Dressler, A., 1980, {\it ApJ}, 236, 351

\bibitem{Dr09}Driver, S.~P., Norberg, P., Baldry, I.~K., et al.\ 2009, Astronomy and Geophysics, 50, 5.12

\bibitem{Forb06}Forbes, D.~A., Ponman, T., Pearce, F., et al.\ 2006, {\it PASA}, 23, 38

\bibitem{Hay11}Haynes, M.~P., Giovanelli, R., Martin, A.~M., et al.\ 2011, {\it AJ}, 142, 170

\bibitem{H17}Hess, K.~M., Cluver, M.~E., Yahya, S., et al.\ 2017, {\it MNRAS} , 464, 957

\bibitem{H13}Hess, K.~M., \& Wilcots, E.~M.\ 2013, {\it AJ}, 146, 124

\bibitem{Leis16} Leisman, L., Haynes, M.~P., Giovanelli, R., et al.\ 2016, {\it MNRAS}, 463, 1692

\bibitem{Jarv12} Jarvis, M.~J.\ 2012, African Skies, 16, 44

\bibitem{Nul82}Nulsen, P.E.J., 1982, {\it MNRAS}, 198, 1007

\bibitem{Os11}Ostriker, E.C., Shetty, R., 2011, {\it ApJ}, 731, 41

\bibitem{Peng10} Peng, Y.-j., Lilly, S.~J., Kova{\v c}, K., et al.\ 2010, {\it ApJ}, 721, 193 

\bibitem{Ras12}Rasmussen, J. et al. 2012, {\it ApJ}, 747, 31

\bibitem{Ras08}Rasmussen, J., et al. 2008, {\it MNRAS}, 388, 1245

\bibitem{Rob13}Robotham, A.~S.~G., Liske, J., Driver, S.~P., et al.\ 2013, {\it MNRAS}, 431, 167

\bibitem{Ser13} Serra, P., Koribalski, B., Duc, P.-A., et al.\ 2013, {\it MNRAS}, 428, 370

\bibitem{Ser11}Serra, P.\ 2011, Fornax, Virgo, Coma et al., Stellar Systems in High Density Environments, 49 



\bibitem{Sch14} Schawinski, K., Urry, C.~M., Simmons, B.~D., et al.\ 2014, {\it MNRAS}, 440, 889
 
\bibitem{Sch07} Schiminovich, D., Wyder, T.~K., Martin, D.~C., et al.\ 2007, {\it ApJS}, 173, 315 
 
\bibitem{Str01} Strateva, I., Ivezi{\'c}, {\v Z}., Knapp, G.~R., et al.\ 2001,  {\it AJ}, 122, 1861
 
\bibitem{TT72}Toomre, A., \& Toomre, J.\ 1972, {\it ApJ}, 178, 623

\bibitem{Ver01}Verdes-Montenegro, L., Yun, M.~S., Williams, B.~A., et al.\ 2001, {\it A\& A}, 377, 812

\bibitem{Vog16} Vogt, F.~P.~A., Owen, C.~I., Verdes-Montenegro, L., \& Borthakur, S.\ 2016, {\it ApJ}, 818, 115 

\bibitem{Vog15} Vogt, F.~P.~A., Dopita, M.~A., Borthakur, S., et al.\ 2015, {\it MNRAS}, 450, 2593 



\bibitem{Yun94}Yun, M.~S., Ho, P.~T.~P., \& Lo, K.~Y.\ 1994, {\it Nature}, 372, 530
 
 
\end{thebibliography}
\end{document}